\documentclass[aps,prl,twocolumn,superscriptaddress,showpacs]{revtex4}
\usepackage{graphicx}
\begin{document}


\title{Particle production at very low transverse momenta \\
in Au+Au collisions 
at $\sqrt{s_{_{NN}}} =$ 200~GeV}


\author{
%
%
B.B.Back$^1$,
M.D.Baker$^2$,
M.Ballintijn$^4$,
D.S.Barton$^2$,
R.R.Betts$^6$,
A.A.Bickley$^7$,
R.Bindel$^7$,
A.Budzanowski$^3$,
W.Busza$^4$,
A.Carroll$^2$,
M.P.Decowski$^4$,
E.Garc\'{\i}a$^6$,
N.George$^{1,2}$,
K.Gulbrandsen$^4$,
S.Gushue$^2$,
C.Halliwell$^6$,
J.Hamblen$^8$,
G.A.Heintzelman$^2$,
C.Henderson$^4$,
D.J.Hofman$^6$,
R.S.Hollis$^6$,
R.Ho\l y\'{n}ski$^3$,
B.Holzman$^2$,
A.Iordanova$^6$,
E.Johnson$^8$,
J.L.Kane$^4$,
J.Katzy$^{4,6}$,
N.Khan$^8$,
W.Kucewicz$^6$,
P.Kulinich$^4$,
C.M.Kuo$^5$,
W.T.Lin$^5$,
S.Manly$^8$,
D.McLeod$^6$,
A.C.Mignerey$^7$,
R.Nouicer$^6$,
A.Olszewski$^3$,
R.Pak$^2$,
I.C.Park$^8$,
H.Pernegger$^4$,
C.Reed$^4$,
L.P.Remsberg$^2$,
M.Reuter$^6$,
C.Roland$^4$,
G.Roland$^4$,
L.Rosenberg$^4$,
J.Sagerer$^6$,
P.Sarin$^4$,
P.Sawicki$^3$,
W.Skulski$^8$,
S.G.Steadman$^4$,
P.Steinberg$^2$,
G.S.F.Stephans$^4$,
A.Sukhanov$^2$,
J.-L.Tang$^5$,
A.Trzupek$^3$,
C.Vale$^4$,
G.J.van~Nieuwenhuizen$^4$,
R.Verdier$^4$,
F.L.H.Wolfs$^8$,
B.Wosiek$^3$,
K.Wo\'{z}niak$^3$,
A.H.Wuosmaa$^1$,
B.Wys\l ouch$^4$\\
\vspace{3mm}
\small
%
%
%
%
$^1$~Argonne National Laboratory, Argonne, IL 60439-4843, USA\\
$^2$~Brookhaven National Laboratory, Upton, NY 11973-5000, USA\\
$^3$~Institute of Nuclear Physics PAS, Krak\'{o}w, Poland\\
$^4$~Massachusetts Institute of Technology, Cambridge, MA 02139-4307, USA\\
$^5$~National Central University, Chung-Li, Taiwan\\
$^6$~University of Illinois at Chicago, Chicago, IL 60607-7059, USA\\
$^7$~University of Maryland, College Park, MD 20742, USA\\
$^8$~University of Rochester, Rochester, NY 14627, USA\\
}

\noaffiliation

\date{\today}

\begin{abstract}
We present results on charged particle production
at very low transverse momenta in the 15\% most central Au+Au collisions 
at $\sqrt{s_{_{NN}}} =$ 200~GeV obtained with the PHOBOS detector at RHIC. The
invariant yields were measured at mid-rapidity in the transverse
momentum ranges from 30 to 50 MeV/c for charged pions, 90 to 130 MeV/c for
charged kaons and 140 to 210 MeV/c for protons and antiprotons. No significant
enhancement in low transverse momentum particle production is observed as 
compared
to extrapolations of identified particle spectra measured at an intermediate 
$p_T$ range.
The spectra tend to flatten at low $p_T$, consistent with the expectations of
transverse expansion of the system.
\end{abstract}


\maketitle

Collisions of gold nuclei at the Relativistic Heavy Ion Collider 
(RHIC)
provide the means to study 
strongly interacting matter under conditions of high temperature and energy
density. 
The study of low-$p_T$ particle production is particularly 
interesting as it is directly 
associated
with the long-distance scales, accessible in heavy ion collisions but out 
of reach
in hadronic interactions. 
Any enhancement of the low-$p_T$ yields compared to extrapolations
from higher transverse momenta could indicate interesting effects, e.g.
new long-wavelength
physics phenomena \cite{buszalowpt}.
Pion production may also be modified if 
a transient state with partially
restored chiral symmetry is produced in the early stage of the collision 
\cite{blaizot,bjorken,Randrup}. 
 
The spectra of  
identified hadrons measured at RHIC at $\sqrt{s_{_{NN}}} =$ 200~GeV 
\cite{stamidpt,phemidpt,bramidpt} 
tend to flatten in the low $p_T$ region. This effect is more pronounced 
for heavier
particles and is commonly 
attributed to a collective
transverse expansion of the system. Hydrodynamics-based models incorporating
 transverse
expansion provide a satisfactory description of the spectra measured above 
0.2 GeV/c 
\cite{hydro1,hydro2,hydro3}. 
Other mechanisms, such as initial state parton interactions \cite{pajares,
saturat} 
or final state 
hadron reinteractions, can also describe the
broadening of the spectra
and an increase of $\langle p_T \rangle$ with particle mass. 
Still, if transverse flow develops during the evolution of the system produced 
in heavy ion
collisions, the effects of flow would likely be observed at 
very low 
$p_T$. 

In this paper we present data on particle production at very low transverse
momenta, $p_T < 0.2$ GeV/c, a region thus far not explored at RHIC energies. 
The PHOBOS detector at RHIC has the unique capability
to detect particles with very low $p_T$ 
in the two-arm spectrometer \cite{phodet}. 
Each spectrometer arm consists of 16 layers of 
Si detectors.
The six innermost layers are located 
in an approximately field-free region near the beam pipe. 
The proximity of the sensitive detector layers
to the interaction region, the small amount of material between the collision 
vertex and the Si
layers, and the high segmentation of the Si detectors permit a precise 
measurement of
particles with small $p_T$. 

The data were taken 
during the 2001 RHIC run. 
The event triggering, determination of
the collision centrality and the position of the collision vertex were as
described in \cite{phodet,phocent}. For this 
analysis
we selected the 15\% most central Au+Au collisions ($\langle N_{part} \rangle$
= 303, where $N_{part}$ is the number of participating nucleons).
Events with the 
reconstructed
vertex position along the beam axis between -7 cm and +14 cm, 
relative to the nominal interaction
point were accepted. The centrality and vertex selections yielded 
$2\times
10^6$ events.

\begin{figure}
\includegraphics[width=6.9cm]{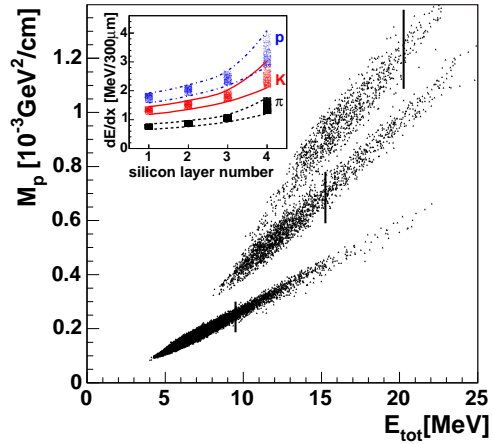}
\caption{\label{fig:fig1} $(M_p, E_{tot}$) scatter plot for the candidate 
tracks after the
$(dE/dx)_i$ cuts in the first four spectrometer layers. 
Only track candidates with emission angles 
larger than $60^o$ are shown. The upper limits on 
$E_{tot}$ 
are marked by vertical lines. 
An inset shows $1\sigma$ bands on the specific energy loss in the
first four Si layers for simulated particles.}
\end{figure}

In this analysis we searched for particles stopped in the $5^{th}$ 
spectrometer
layer. 
The reconstruction of particle trajectories and determination of the
particle mass was based on a detailed analysis of the measured energy
depositions in the first five spectrometer layers. Selection criteria based on
the energy depositions \cite{phosil} for track candidates were 
derived from an
analysis of single stopping particles obtained from GEANT
simulations of the PHOBOS detector \cite{geant}. Initially, all possible
5-hit track candidates were formed from hits with 
large energy 
depositions
($dE/dx > 0.5$ MeV normalized to 300$\mu m$ of Si, 
roughly 6 times the deposition of a minimum ionizing
particle). 
For the kinetic
energies of particles stopping in the $5^{th}$ layer (about 8 MeV 
for pions, 
19 MeV for
kaons and 21 MeV for protons) the energy losses depend on the particle
mass \cite{pdg,leo} as shown in the inset in Fig.~\ref{fig:fig1}.
Each track candidate satisfied the requirement $|(dE/dx)_i - <dE/dx>_i| < 
1.4 \sigma_i$, $i=1,...,4$,  where the mean and the rms value $\sigma_i$, 
are calculated from 
the $(dE/dx)_i$ 
distributions
for single simulated stopping particles.

A mass
parameter was defined as 
$(M_p)_i = (dE/dx)_i \sum_{k=i}^{5}E_k$, for the first four layers, 
where $E_k$ denotes the energy
deposited in the $k^{th}$ layer. For non-relativistic particles, the mass
parameter $(M_p)_i$ depends only on the particle mass, since 
$(dE/dx)_i \sim 1/\beta^2$ and $\sum_{k=i}^{5}E_k \sim m\beta^2/2$.
Since the measured energy deposition
in the fifth layer ($E_5$)
cannot be used, due to contributions from annihilation 
and decay products, we estimated $E_5$ using the properties of the 
mass parameter and the assumption that the particle stops in 
the $5^{th}$ layer. 
Having estimated $E_5$ we calculate the average mass parameter,
$M_p$, and the total deposited energy,
$E_{tot} = \sum_{i=1}^{5} E_i$.  
Fig.~\ref{fig:fig1} shows a scatter plot of $M_p$ vs. $E_{tot}$, 
for track candidates found in the data which satisfy 
the selection criteria  on $dE/dx$ in 
the first four
Si layers. 
 
The momenta of 
stopped particles are restricted by imposing cuts on $E_{tot}$
(see  Fig.~\ref{fig:fig1}). For track candidates with emission angles
smaller than $60^o$, not shown in Fig.~\ref{fig:fig1}, about 10\% higher
upper limits on $E_{tot}$ were applied as compared to those shown in 
Fig.~\ref{fig:fig1}.
Additional cuts on the angular
deviations of the candidate track from a straight-line trajectory were
imposed to reject false tracks. The cuts are mass dependent
to account for differences in the multiple scattering of low-momentum 
pions, kaons and protons. This analysis procedure was tested 
on samples of simulated low-momentum charged particles. 
Reconstructed single tracks
simulated by GEANT were used to calibrate the final particle momenta.
The transverse momentum resolution is estimated to be approximately
5\%. These measurements are confined to the mid-rapidity region 
($-0.1 < y < 0.4$) and
cover a limited range of transverse momenta: 30 to 50 MeV/c for charged pions, 
90 to 130 MeV/c for
charged kaons and 140 to 210 MeV/c for protons and antiprotons.

For further analysis the reconstructed particles were divided  into
narrow ($\Delta y, \Delta p_T$) bins. For each bin, corrections were applied
to the raw yields to account for  
acceptance, reconstruction and particle identification 
inefficiencies,
and absorption in the beam pipe. These corrections were obtained by embedding 
simulated low-momentum 
particle tracks into real data events, and range between 
$0.2$ and $0.7$\%. 

\begin{figure}
\includegraphics[width=6.7cm]{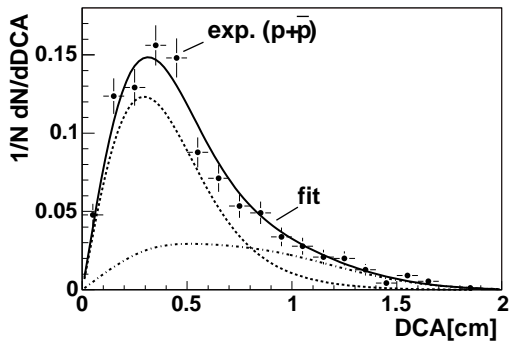}
\caption{\label{fig:fig3} $DCA$ distribution for reconstructed proton tracks 
in the
data sample (points with
error bars).
The result of the fit to the measured distribution (see text for more 
explanations)
is denoted by the solid curve. The dashed (dashed-dotted) curve shows the 
contribution
from primary (background) particles.}
\end{figure}

The acceptance and efficiency corrected particle yields are contaminated
by various background contributions: feed-down from weak decays and 
contributions
from secondary, misidentified and ghost particles. 
For protons and antiprotons the
dominant background contributions originate from secondary particles produced 
in the beam pipe or detector material and 
products 
from weak decays. To estimate the background contamination of the 
($p + \overline{p}$) yields, 
we analyzed the distributions 
of the measured distance of closest approach  ($DCA$) from the vertex position 
to the particle
trajectory, determined by a straight line passing through the hits in the
first and third spectrometer layers. 
The experimental $DCA$ distribution  
is broader and has a 
tail to larger $DCA$ values compared to the same distribution for 
reconstructed
primary proton tracks. Parameterized shapes of the
$DCA$ distributions for reconstructed primary, secondary and feed-down
protons, obtained from an analysis of HIJING events \cite{hijing},
were used to describe the measured $DCA$ spectrum. 
An example of the fit, assuming equal
weights of the two background components, 
is shown in Fig.~\ref{fig:fig3}. The fitted fraction of background
particles, averaged over all possible weights for the two background
components amounts to $0.31 \pm 0.04(stat.) \pm 0.19(syst.)$. The large
systematic error assigned to this fraction  reflects the dependence of 
the total background estimate
on the relative importance of secondary and feed-down contributions. The
fraction of background particles thus determined agrees with the estimate of
$0.34 \pm 0.03$, obtained directly from the reconstruction of HIJING events.


To estimate the background corrections for kaons and pions we used HIJING 
simulations with 
yields tuned to account for the fact that HIJING 
overpredicts the 
yields of different particle species \cite{lowptqm}. 
This adjustment of HIJING yields is important
for evaluating the background correction for charged kaons which is dominated
by misidentified protons. The estimated background in the measured kaon sample 
is 
$0.10 \pm 0.04$, while the background level in the unscaled HIJING events
is higher ($0.23 \pm 0.03$) due to the larger relative yield of protons to 
kaons. For protons and kaons  the corrections are averaged over 
the full $p_T$ range.

The charged pion background estimates for
scaled and unscaled HIJING events
agree. In this case, feed-down from weak decays and ghost tracks are the
main source of 
the background contamination. The large number of reconstructed pions 
in the
simulated events permit the determination of the $p_T$ dependence of the 
background 
corrections,
which varies between $0.31 \pm 0.04$ and $0.22 \pm 0.02$ for $p_T$ of 
0.03~GeV/c 
to 0.05~GeV/c.

The relative statistical uncertainties from the data 
as well as the statistical errors on the acceptance, efficiency and background
corrections are 4 \% for $\pi^{\pm}$, 10 \% for $K^{\pm}$ and 
12 \% for
$(p+\overline{p})$.
Various sources of systematic errors were investigated, related to
the backgrounds, the analysis method, and detector effects.
Table~\ref{table:table2} summarizes the final estimates of the systematic
uncertainties. 

The dominant contribution to the systematic errors is due to the uncertainty in
the background estimates which are evaluated to be approximately 28\% 
for $(p+\overline{p})$
and about 10\% for charged pions and kaons.
 
The uncertainties related to the analysis method were estimated
by performing the reconstruction with varied selection criteria. There
are also contributions estimated by the 
reconstruction of
HIJING events, and of events which were generated with $p_T$ spectra 
approximately consistent with
our measured data and with intermediat- $p_T$ results
from other experiments \cite{stamidpt,phemidpt,bramidpt}. 
In addition, a 5\% uncertainty in the $p_T$ 
scale
and the sensitivity to the GEANT energy threshold contribute to the
systematic uncertainty of the reconstruction method.
 
The dominant contributions to the detector-related error come
 from the differences in 
particle yields measured separately in the two spectrometer 
arms and under
different magnetic field polarities, and the calibration of energy depositions.
 Other detector effects such as 
misalignment of the spectrometer layers, and the 
accuracy
of the vertex position, were found to be negligible.

The individual contributions from different sources of systematic
uncertainty are added in quadrature to derive the total systematic error
for these measurements: 
$12$\% for $\pi^{\pm}$, 
18\% for $K^{\pm}$ and 32\% for
$(p+\overline{p})$.
\begin{table}
\caption{\label{table:table2} The estimated systematic errors (\%) on the measured
invariant yields.}
\begin{ruledtabular}
\begin{tabular}{crrr}
Source & $\pi^+ + \pi^-$ & $K^+ + K^-$ & $p+\overline{p}$ \\ \hline
Method & 7.5  & 12.1  & 13.5  \\ 
Detector & 3.8  & 7.7  & 9.9 \\ 
Background & 9.0  & 10.0  & 27.5 \\ 
\end{tabular}
\end{ruledtabular}
\end{table}
 

%

\begin{figure}[th]
\includegraphics[width=6.4cm]{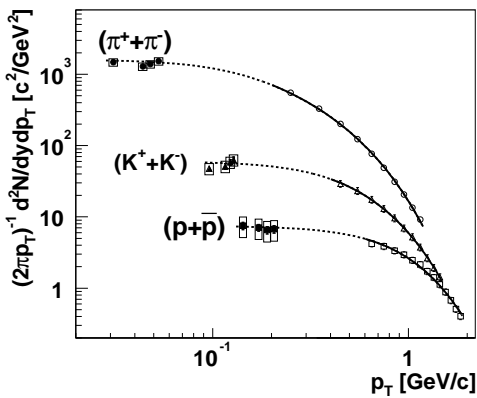}
\caption{\label{fig:fig6} Invariant yields as a function of $p_T$. 
For low-$p_T$
yields (closed symbols) the boxes show systematic uncertainties. 
For comparison the intermediate-$p_T$ measurements from PHENIX \cite{phemidpt} 
are depicted by open
symbols. The fits to PHENIX measurements 
(solid curves) are extrapolated
to low $p_T$ (dashed curves). See text for more details.}
\end{figure}

The fully corrected invariant yields, $(2\pi p_T)^{-1} d^2N/dp_T dy$,
of charged particles produced in the 
15\% most central Au+Au collisions at $\sqrt{s_{_{NN}}} =$ 200~GeV
are shown in Fig.~\ref{fig:fig6}, together with the yields of 
identified hadrons measured by PHENIX 
\cite{phemidpt} for the same centrality selection. Different parameterizations 
have been
used  to extrapolate the measurements at higher $p_T$
to $p_T \approx 0$ in order to determine the mean $p_T$ and integrated 
particle yields 
\cite{stamidpt,phemidpt,bramidpt}. 
Measurements at very low $p_T$ provide some constraints on the choice of the functional
form which fits the $p_T$ spectra. The functional form 
$A [exp(m_T/T_{fit})+
\epsilon]^{-1}$, with $\epsilon = -1 (+1)$ for mesons (baryons), 
reasonably well
describes the intermediate-$p_T$ PHENIX measurements ($m_T \leq 1~GeV/c^2$).
The fit parameter,
$T_{fit}$, equals $0.229 \pm 0.005$, 
$0.293 \pm 0.01$ and
$0.392 \pm 0.015~GeV/c^2$ correspondingly for charged pions, kaons and 
$(p+\overline{p})$.
An 
extrapolation of the fits to low $p_T$ agrees with the measured 
low-$p_T$ yields 
(see Fig.~\ref{fig:fig6}). 

\begin{figure}[th]
\includegraphics[width=6.8cm]{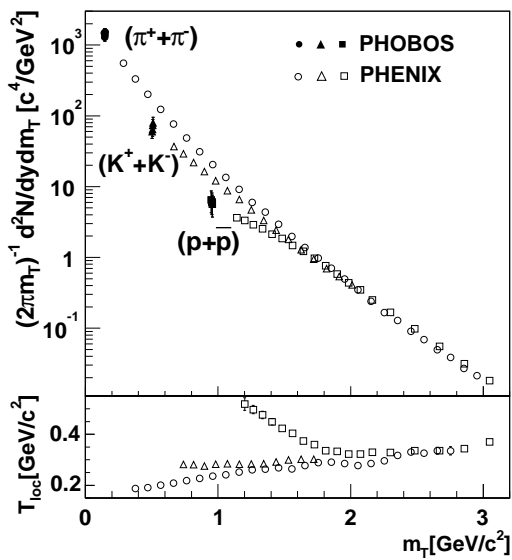}
\caption{\label{fig:fig7} Invariant yields, normalized at $m_T = 2 GeV/c^2$, 
as a function of $m_T$. Lower panel shows the $m_T$ dependence of the local 
inverse 
slopes.}
\end{figure} 

It was predicted  that the $m_T$ spectra of different particle species
should follow a universal function of  $m_T$ \cite{pajares,saturat}.
One implication of the  $m_T$ scaling described in \cite{saturat} is that the
spectral shapes should be independent of particle mass over a wide range 
of $m_T$. 
In Fig.~\ref{fig:fig7} the yields normalized at $m_T= 2~ GeV/c^2$ are
shown as a function of $m_T$. This choice of normalization was motivated by
an analysis of the 
local
inverse slopes, $T_{loc}$, obtained from the exponential fits done with
a sliding window enclosing 5 consecutive $m_T$ points. (The first inverse slope
is from the fit to our  average yield and the first four PHENIX points.) 
As illustrated in the lower panel of Fig.~\ref{fig:fig7}, these local inverse 
slopes converge at $m_T \approx 2~ GeV/c^2$.  
The normalized $m_T$ spectra are consistent with the scaling hypothesis 
in the range of $m_T$ from about 1.5 to 2.2 $GeV/c^2$. However at low $m_T$, 
close to
the threshold, the spectra for heavier particles deviate
from the charged pion spectrum, with larger deviations observed for the
$(p+\overline{p})$ yields than for the charged kaons. 
This observation is consistent with 
mass dependent flattening of the spectra at low $m_T$, expected in the
presence of transverse expansion of the system, and contradicts
the suggested $m_T$ scaling \cite{saturat}.

In summary, the yields of identified particles have
been measured at very low $p_T$ (below 210 MeV/c) in 
central $\sqrt{s_{_{NN}}} =$ 200~GeV Au+Au collisions. 
We see no enhancement 
in the production of
particles with very low transverse momenta, which might have indicated
 the presence
of unusual long-wavelength phenomena. The low-$p_T$ yields agree
with extrapolations from intermediate-$p_T$ measurements. 
Finally, the measurements in this region of phase space, unexplored until now,
show clear flattening of particle spectra that 
increases
with particle mass. This observation supports an interpretation
of the data in terms of a collective transverse expansion
of the system.

\begin{acknowledgments}
This work was partially supported by U.S. DOE grants DE-AC02-98CH10886,
DE-FG02-93ER40802, DE-FC02-94ER40818, DE-FG02-94ER40865, DE-FG02-99ER41099, and
W-31-109-ENG-38 as well as NSF grants 9603486, 9722606 and 0072204.  The Polish
group was partially supported by KBN grant 2-P03B-10323.  The NCU group was
partially supported by NSC of Taiwan under contract NSC 89-2112-M-008-024.
\end{acknowledgments}


\end{document}